\documentclass[pdflatex,sn-mathphys]{sn-jnl}% Math and Physical Sciences Reference Style

\jyear{2021}%

\theoremstyle{thmstyleone}%
\theoremstyle{thmstyletwo}%

\theoremstyle{thmstylethree}%

\begin{document}

\title[Comparison of Home Detection Algorithms using Smartphone GPS Data]{Comparison of Home Detection Algorithms using Smartphone GPS Data}
    
\author[1]{\fnm{Rajat} \sur{Verma}}
\author[1]{\fnm{Shagun} \sur{Mittal}}
\author[1]{\fnm{Zengxiang} \sur{Lei}}
\author[1]{\fnm{Xiaowei} \sur{Chen}}

\author*[1]{\fnm{Satish V.} \sur{Ukkusuri}}\email{sukkusur@purdue.edu}

\affil*[1]{\orgdiv{Lyles School of Civil Engineering}, \orgname{Purdue University}, \orgaddress{\street{550 Stadium Mall Avenue}, \city{West Lafayette}, \postcode{47907}, \state{Indiana}, \country{USA}}}

\abstract{
% Background
Estimation of people's home locations using location-based services data from smartphones is a common task in human mobility assessment. However, commonly used home detection algorithms (HDAs) are often arbitrary and unexamined.
In this study, we review existing HDAs and examine five HDAs using eight high-quality mobile phone geolocation datasets.
These include four commonly used HDAs as well as an HDA proposed in this work.
To make quantitative comparisons, we propose three novel metrics to assess the quality of detected home locations and test them on eight datasets across four U.S. cities.
We find that all three metrics show a consistent rank of HDAs' performances, with the proposed HDA outperforming the others. We infer that the temporal and spatial continuity of the geolocation data points matters more than the overall size of the data for accurate home detection.
We also find that HDAs with high (and similar) performance metrics tend to create results with better consistency and closer to common expectations. Further, the performance deteriorates with decreasing data quality of the devices, though the patterns of relative performance persist.
Finally, we show how the differences in home detection can lead to substantial differences in subsequent inferences using two case studies – (i) hurricane evacuation estimation, and (ii) correlation of mobility patterns with socioeconomic status.
Our work contributes to improving the transparency of large-scale human mobility assessment applications.
}


\maketitle

\section{Introduction}\label{sec:intro}
% what is home detection
% Human mobility assessment (HMA) is an important task in fields such as business placement, urban planning, transportation and traffic safety engineering, congestion and hotspot analysis, behavioral studies, migration studies, and equity analysis \cite{barbosa2018human,yabe2020quantiyfing,wang2019urban}. HMA involves tasks such as identifying spatiotemporal patterns of movements such as general trip making \cite{Liu2017}, commuting \cite{kung2014exploring}, evacuation and displacement \cite{yabe2019cross}, and migration \cite{luca2021leveraging}. 

% Among these tasks, estimation of people's home location is a necessary step in many subsequent tasks. It is important to correct data representativeness bias (for example, identifying areas that are under or overrepresented in the data) \cite{Shen2014,guo2020systematic}, understand socioeconomic differences in mobility during regular and hazardous situations \cite{verma2021spatiotemporal}, perform urban planning considering accessibility and equity, and augment travel surveys \cite{guo2020systematic}.

Home location detection is an important step in several fields of human mobility analysis such as transportation planning \cite{ccolak2015analyzing}, migration and evacuation studies \cite{lai2019exploring,yabe2022mobile}, accessibility analysis \cite{guo2020systematic}, and the theory of human mobility \cite{schneider2013unravelling,gonzalez2008understanding}. This task involves predicting people's `home location' based on geolocation data, often collected passively by their devices via location-based services, call detailed records, social media activity, smart-card transactions, and in-vehicle location trackers \cite{anda2017transport}.

% why home detection is needed
Home detection plays an essential role in understanding large-scale human mobility patterns. For instance, in the event of a hurricane, one needs the home locations both before and after the disaster to identify their evacuation status \cite{yabe2019cross}. In urban planning, identifying home locations serves as the foundational data for vital information including home-based trips \cite{demissie2018trip} and human mobility metrics \cite{song2010modelling}, and this forms the basis for optimizing existing infrastructure \cite{harrison2020new}. It is consequently important to have a robust understanding of home detection approaches.

% problem 1: lack of systematic review of HDAs
Despite its significance, existing studies using home detection algorithms (HDAs) have paid little attention to the effectiveness of their algorithms. Researchers have developed several HDAs for geolocation data of different kinds whose assumptions, methods, and parameters are not necessarily consistent with one another \cite{pappalardo2021evaluation}. This raises doubts about the validity of their findings as the error in home detection may propagate to the downstream calculation of home-related metrics such as evacuation counts \cite{yabe2022mobile}, home-based trip rates \cite{dypvik2021mobile}, and data representativeness figures for accessibility analysis \cite{garcia2019exploring}.

% problem 2: lack of supervised method
This issue stems primarily from a lack of ground truth home locations associated with large geolocation datasets. The collection of accurate home location collection on a large scale poses significant risks to privacy \cite{jiang2013review}. Mobility data vendors provide anonymized device identifiers and modify sensitive trajectories to prevent an accurate tracking of people's trip origins and destinations \cite{yabe2022mobile}. In the absence of ground truth data, it becomes difficult to compare the accuracy of different HDAs using supervised learning methods. Researchers have largely relied on unsupervised methods for home detection, such as rule- and clustering-based HDAs.
Small-scale studies such as \cite{pappalardo2021evaluation} and \cite{vanhoof2018assessing} have sought to compare the effectiveness of HDAs but have focused only on the parameters of a few HDAs. Further, their small experiments do not provide insights about the impact of study region and period and data quality on the performance of the HDAs.

% Contributions
In this study, we tackle this issue of a lack of a systematic comparative assessment of commonly used HDAs. In doing so, we contribute to the literature on the home location detection problem in the following ways:
\begin{enumerate}
    \item We review the state-of-the-art HDAs that use large-scale mobility data, including their benefits, assumptions, and limitations.
   \item We propose three intuitive metrics to quantify the quality of the home-location detection results in the absence of ground truth home location information.
    \item We develop a comprehensive experiment where a set of HDAs are quantitatively compared in terms of the introduced performance metrics and their sensitivity to the data quality.
    \item We propose a new HDA that overcomes some of the limitations of the above methods and shows superior performance.
\end{enumerate}

\begin{figure}
    \centering
    \includegraphics[width=\textwidth]{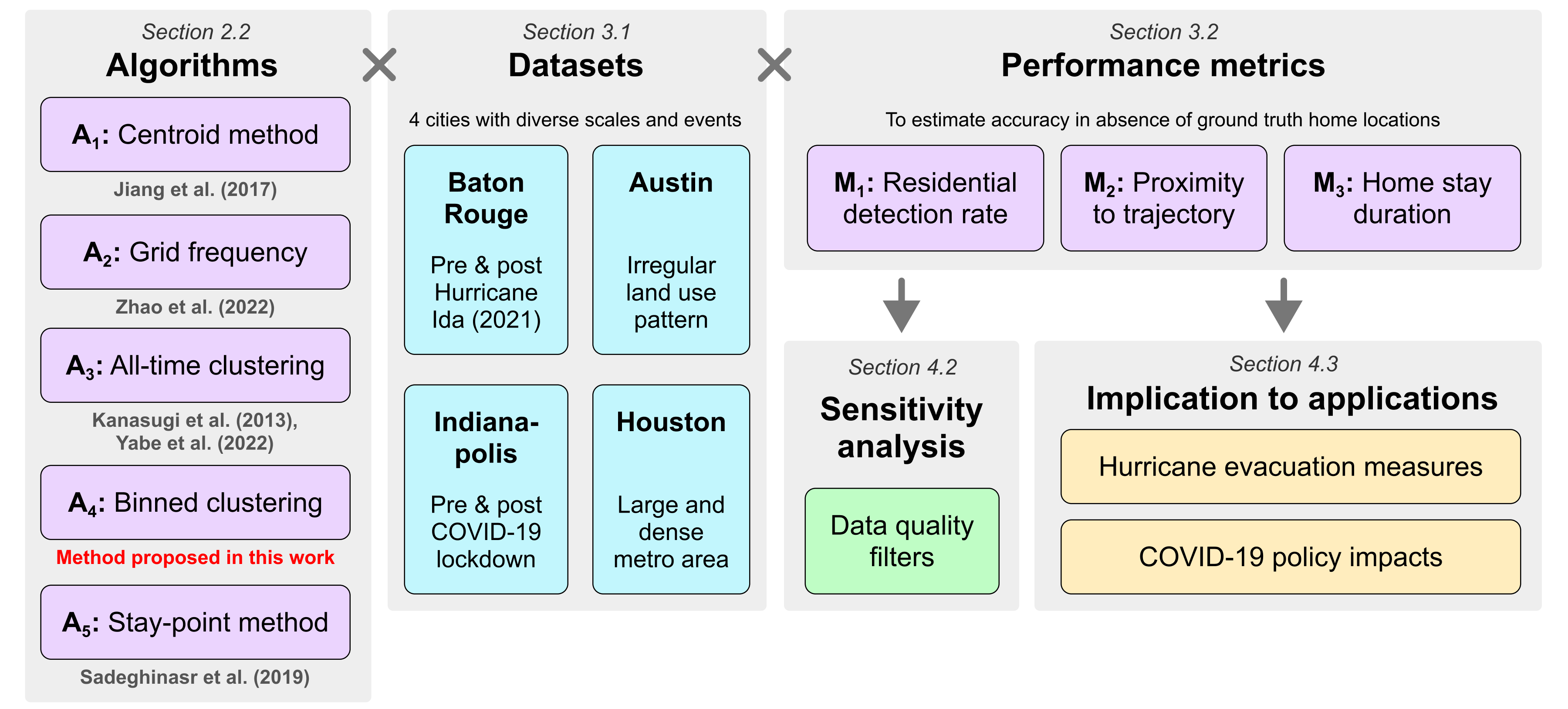}
    \caption{\textbf{Framework of the study}. The figure shows the key components of the experiment – HDAs, datasets, and metrics. The cross symbol denotes Cartesian product.}
    \label{fig:overall-framework}
\end{figure}

The framework and experiment design of this study is shown in Fig.~\ref{fig:overall-framework} and described in detail in the following sections.
The main objective is to compare the performance of different HDAs across different input datasets. On the basis of the review of research literature on HDAs, we have selected four popular and unique HDAs, and additionally proposed an HDA for comparison in this study.
The testing is done on eight input samples of passively collected smartphone GPS data covering four U.S. metropolitan areas of different data qualities and different time periods spawned by mobility-influencing events.
% To this end, different input samples are generated for different combinations of study regions and time periods.
%and data filtering criteria such as the nighttime period and the minimum ping count as a data quality control parameter.
Once home locations are estimated for each combination of the sample dataset and HDA, they are compared using three approximate accuracy metrics proposed in Section \ref{sec:metrics}. The performance of the HDAs under the different dataset conditions is discussed in Section \ref{sec:performance-comparison}. In the subsequent sensitivity analysis Section \ref{sec:sensitivity}, the performance metrics are recomputed for different subsamples of the datasets by changing the quality of the users in the input dataset. Finally, the impacts of these HDAs on subsequent applications, such as hurricane evacuation assessment and analysis of mobility change during COVID-19, are shown in Section \ref{sec:impact-to-apps}.

% \section{Related work}
\section{Home detection algorithms}

\subsection{Literature review}
% \subsubsection{Mobility data types}

% introduction to lit review table
HDAs from mobility data can be categorized on the basis of several characteristics, such as the type of input geolocation data (such as social media and passively collected GPS data), the modeling paradigm (supervised vs.\ unsupervised and rule-based vs.\ data-driven), and constraints for filtering the input data. Based on these classifications, some prominent HDAs are reviewed and summarized in Table \ref{tab:lit-review}.
% This section introduces kinds of related research by supervised versus unsupervised models. 
% Supervised HDAs involve using user-verified home locations to train machine learning models on mobility data, whereas unsupervised HDAs rely on intuitive assumptions related to the patterns of homes.
% supervised vs unsupervised
% Based on the data type and attributes available to the researcher, HDAs can be broadly categorized into two kinds – supervised and unsupervised.
% HDAs from mobility data and their outcomes are substantially influenced by the type, quality, and quantity of input data \cite{Shen2014}.
% Major mobility data types used in most home detection applications include GPS travel surveys, social media location data, smartcard transactions, call detailed records (CDR), and passively collected GPS data from mobile devices and automobiles.

% supervised - GPS surveys: what, who used it
\subsubsection{Supervised methods}
Supervised methods predominantly rely on GPS-based travel surveys that involve the subjects carrying GPS-enabled devices that track movements. In addition to individual-level information such as actual (ground truth) home locations, demographic characteristics, and personal preferences, the devices provide detailed travel entries such as the origin and destination, the departure and arrival time, the trip purpose, and the travel mode. Such mixed methods have been used in many pilot studies \cite{zhou2007discovering,bayat2020inferring,pappalardo2021evaluation}.
In some cases, it is also possible to obtain CDR data of specific groups for whom individual-level data is also available, such as employees of a telephone carrier \cite{pappalardo2021evaluation}.
% travel diaries
In other cases, such as in standalone travel diaries like the National Household Travel Survey, the respondents' street addresses are geocoded to coordinates, though other large-scale information is not obtained for them \cite{jenkins2023}.
% supervised - models
With true home locations of the small survey sample, it is possible to create sophisticated supervised machine learning models, such as random forests and AdaBoost \cite{oosterlinck2021home} or artificial neural networks \cite{leca2015significant}.

% supervised - limitations
Although supervised HDAs are powerful, they suffer from a major limitation of training data availability due to privacy reasons. In the recent past, growing pressure from human rights organizations and the subsequent government regulations has made it difficult to obtain actual home locations of individuals at a large scale \cite{stalla2016anonymous}. GPS surveys and CDR samples used in supervised HDAs are usually very small, often with fewer than 100 subjects \cite{bayat2020inferring,pappalardo2021evaluation}. Samples also typically form specialized volunteer subjects such as students \cite{isaacman2011identifying} and older patients \cite{wojtusiak2021location}, raising concerns about sample representativeness \cite{vanhoof2018assessing}. In addition, GPS travel surveys do not represent longitudinal data. These issues make supervised HDAs much less popular in the research literature.

% unsupervised - pros (large, cheap, temporal continuity) & cons (no demographics)
\subsubsection{Unsupervised methods}\label{sec:unsupervised-methods}
\textbf{\textit{Basic assumptions}}\\
Due to the difficulty in obtaining high-quality home location data at a large scale, researchers have relied heavily on unsupervised HDAs. These methods necessarily depend on a set of assumptions about people's home locations that are found throughout the research literature \cite{sadilek2012modeling,leca2015significant,yabe2022mobile}. These include:
\begin{itemize}
    \item People are more likely to stay at their homes during the off-work period. This normally includes nighttime, but in some cases, can be extended to weekends or even the office after-hours.
    \item The most observed place for an individual, especially at night, is usually their home.
\end{itemize}
These assumptions intuitively make sense, although there are several exceptions, such as people who work from home or who work night shifts. However, since these assumptions are almost always used in unsupervised HDAs, we consider these assumptions to be axiomatic.

% data types
\vspace{0.2cm} \noindent \textbf{\textit{Dataset types}}\\
The input datasets for unsupervised HDAs are abundantly available on scale, including longitudinal data \cite{luo2016explore,zhao2022estimating}, although they lack the demographics and travel preferences of the subjects \cite{Shen2014,jiang2017activity,sadeghinasr2019estimating}. Some of the most prominent dataset kinds include the following:

\begin{itemize}
% social media data
    \item \underline{Social media location data} include posts on websites such as Twitter, Foursquare, and Flickr that a user tags with the location of the mentioned place \cite{luo2016explore}. They are usually available at large scales and in several time periods but are usually spatiotemporally sparse and biased toward certain demographics for effective home location detection, and access to the data can disappear quickly \cite{jurdak2015understanding,wang2018urban,phillips2021social}.
% smartcard transactions
\item \underline{Smart card data} include transactions at payment booths such as at subway stations and inside public transit buses~\cite{zou2018detecting}. These are usually anonymized and frequent, but they can only be used to infer public transit mobility patterns adequately as opposed to home locations.
% CDR
\item \underline{Call detailed records (CDRs)} provide geolocation data at the cell tower level. Such datasets are characterized by large spatiotemporal density and coverage, but the quality of the detected homes is subject to the spatial distribution of the cell towers rather than the users' activity patterns \cite{xu2015understanding}. Nonetheless, they have been used extensively to understand people’s travel and activity patterns during the recording period \cite{xu2015understanding, jiang2017activity, chen2018enriching}.
% passively collected GPS data
\item \underline{Passively collected GPS data} are usually obtained from mobile devices such as smartphones and tablets and automobiles that have location-based services (LBS) enabled~\cite{sadeghinasr2019estimating}. GPS data overcomes the main problem with CDRs by providing the exact locations and overcomes geotagged posts by providing continuous and high-frequency records. Further, GPS can provide more detailed information about the movements of individuals, including their speed, direction, and stop durations along the way. Therefore, it has seen a substantial increase in availability and use in the last decade. In this study, we use this data kind for our analysis.
% \item For analysts, a major problem with this data kind is the anonymization of the users' identities to protect their privacy.
\end{itemize}

\vspace{0.2cm} \noindent \textbf{\textit{Method Types}}\\
% clustering - k-means and hierarchical
% density-based clustering
\noindent \textbf{Density-based clustering} methods are for home detection (used in \cite{yabe2020effects}).

Mean-shift clustering \cite{com2002} is a popular density-based clustering method that has been used in several studies \cite{yabe2020effects, yabe2022mobile, ashbrook2003using, comito2016mining}, probably owing to its simplicity in having just one main parameter – the radius of flat kernel for kernel density estimation (KDE).
DBSCAN \cite{ester1996density} and its variants (e.g., \cite{shah2012improved,Xu2021}), on the other hand, have two main parameters – the maximum intra-cluster distance at each iteration and the minimum number of points in an acceptable neighborhood.
In both these methods, the results of the clustering can be substantially sensitive to the choice of these parameters \cite{fahad2014survey}.

\noindent \textbf{Heuristic algorithms} are widely applied to detect home locations, which rely on various decision rules on the time and frequencies of user records in specific areas during observations \cite{vanhoof2018assessing,sadeghinasr2019estimating,vanhoof2020performance,pappalardo2021evaluation}. The most intuitive assumption is that users have the highest records at home, and their home locations are identified based on the density of the data. Different variants are proposed by shifting the rules, such as determining their home as locations with the highest number of nighttime records or the most distinct days.

% Stay point detection
Li et al. (2008) \cite{Li2008} developed a rule-based method for detecting `stay points' which represent spatiotemporal regions of low movement and are thus helpful in trip segmentation.
These stay points are computed by identifying the breaks in the time gap and distance between the first and last point of a sequential set of points based on given thresholds of time gap (30 min) and distance (200 m).
This method was further modified by Sadeghinasr et al. (2019) \cite{sadeghinasr2019estimating} who clustered these stay points using hierarchical clustering into stay regions and identified home locations as the most visited stay regions during nighttime.
% Other methods
Other methods, such as the center-point algorithm by Zou et al. (2018)~\cite{zou2018detecting}, which uses one's middle point of the first to-subway trip's origin and the last from-subway trip's destination to represent the home location, are easy to compute but have been shown to perform fairly well up to a large radius of tolerance (e.g., \cite{zou2018detecting,jurdak2015understanding}).

Current studies that compared different HDAs already demonstrated that the results are sensitive to criteria choice, such as night time periods. For instance, Vanhoof et al. \cite{vanhoof2020performance} primarily focused on assessing the effects of different night periods on the home detection results while ignoring the limitations of the HDA. Pappalardo et al. \cite{pappalardo2021evaluation} compared five similar HDAs and validated the results with multiple small-scale datasets, yet they neglected to consider factors such as data quality and period. In contrast, this study concerns comparing the HDAs, with a particular emphasis on testing across scenarios spanning different regions, data periods, and data quality.

\begin{table}[hbt!]
\footnotesize
\centering
\caption{Summary of commonly used HDAs for different data and algorithm types.}
\label{tab:lit-review}
\setlength\tabcolsep{3pt}
\begin{tabu}  to \textwidth {c X[1]  X[1] X[0.6] X[4]}
\toprule
\textbf{Kind}  & \textbf{Algorithm class}  & \textbf{Dataset type} & \textbf{Sources} & \textbf{Definition of home location} \\
\midrule
% ---------------- Supervised --------------------------
\multirow[t]{6}{1.2cm}{Supervised} & 
    \multirow[t]{4}{1.2cm}{Clustering} &
        CDR &
        \cite{isaacman2011identifying} &
        Most popular important cluster (Hartigan clustering of cell towers) using a logistic regression model \\
        \cmidrule(l){3-5}
        &
        &
        GPS Survey / Tracking &
        \cite{bayat2020inferring} &
        Density-based spatial clustering of points with noise (DBSCAN) \\
        \cmidrule(l){4-5}
        &
        &
        &
        \cite{zhou2007discovering} &
        Most popular of the clusters based on DJ-cluster algorithm (modified DBSCAN clustering) \\
        \cmidrule(l){4-5}
        &
        &
        &
        \cite{ashbrook2003using} &
        Most popular of the `locations' (obtained using modified k-means clustering of "places") \\
    \cmidrule(l){2-5}
    &
    Clustering and heuristic &
    CDR &
    \cite{oosterlinck2021home} &
    Binary classification algorithms; logistic regression, random forest, adaboosting and neural network models \\
    \cmidrule(l){2-5}
    &
    Heuristic &
    CDR &
    \cite{vanhoof2018assessing,vanhoof2020performance,pappalardo2021evaluation} &
    Most active tower for several data filter criteria such as nighttime constraints, weekday/weekend, and distinct days \\
    \cmidrule(l){1-5}
    % ---------------- Unsupervised --------------------------
\multirow[t]{14}{1.2cm}{Unsuper-vised} & 
    \multirow[t]{3}{1.2cm}{Clustering} &
    CDR &
    \cite{kanasugi2013spatiotemporal} &
    Most frequent stay place (determined based on mean-shift clustering of sequenced cell tower locations) \\
    \cmidrule(l){3-5}
    &
    &
    \multirow[t]{2}{1.2cm}{Passive GPS} &
    \cite{Sadeghinasr2019} &
    Largest hierarchical cluster of stay points (detected based on Liu et al. (2008)) \\
    \cmidrule(l){4-5}
    &
    &
    &
    \cite{yabe2019mobile,yabe2022mobile} &
    Largest cluster of nighttime records using mean-shift clustering \\
    \cmidrule(l){2-5}
    &
    \multirow[t]{3}{1.2cm}{Heuristic} & 
    \multirow[t]{5}{1.2cm}{CDR} &
    \cite{xu2015understanding} &
    Location of the more popular of the two cell towers with the most records during non-work time \\
    \cmidrule(l){4-5}
    &
    &
    &
    \cite{jiang2017activity} &
    Most frequently communicated tower during nights of weekdays, and weekends over the study period \\
    \cmidrule(l){4-5}
    &
    &
    &
    \cite{chen2018enriching} &
    Most frequent location during night time \\
    \cmidrule(l){4-5}
    &
    &
    &
    \cite{leca2015significant} &
    Most common visited locations during night time \\
    \cmidrule(l){4-5}
    &
    &
    &
    \cite{ahas2010using} &
    Anchor point determination model (cell tower location satisfying specific rules of call count) \\
    \cmidrule(l){3-5}
    &
    &
    Passive GPS &
    \cite{zhao2022estimating} &
    The centroid of the most visited 20 × 20 m cell during night hours \\
    \cmidrule(l){3-5}
    &
    &
    \multirow[t]{5}{1.2cm}{Smart card} &
    \cite{zou2018detecting} &
    Center point-based HDA (iteratively updated centroid between pairs of subway stations) \\
    \cmidrule(l){4-5}
    &
    &
    &
    \cite{hasan2013spatiotemporal} &
    Most visited transit station \\
    \cmidrule(l){4-5}
    &
    &
    &
    \cite{bojic2015choosing} &
    Most popular transaction place (overall and active days); place with most nighttime activity \\
    \cmidrule(l){3-5}
    &
    &
    \multirow[t]{2}{1.2cm}{Social media} &
    \cite{scellato2011socio,cho2011friendship} &
    Place with the most check-ins on 3 social networks \\
    \cmidrule(l){4-5}
    &
    &
    &
    \cite{sadilek2012modeling} &
    Place with the most check-ins during midnight \\
\bottomrule
\end{tabu}
% }
% \end{tabularx}
\end{table}

\subsection{Algorithms used in this study}\label{sec:algorithms}
\begin{figure}
    \centering
    \includegraphics[width=\textwidth]{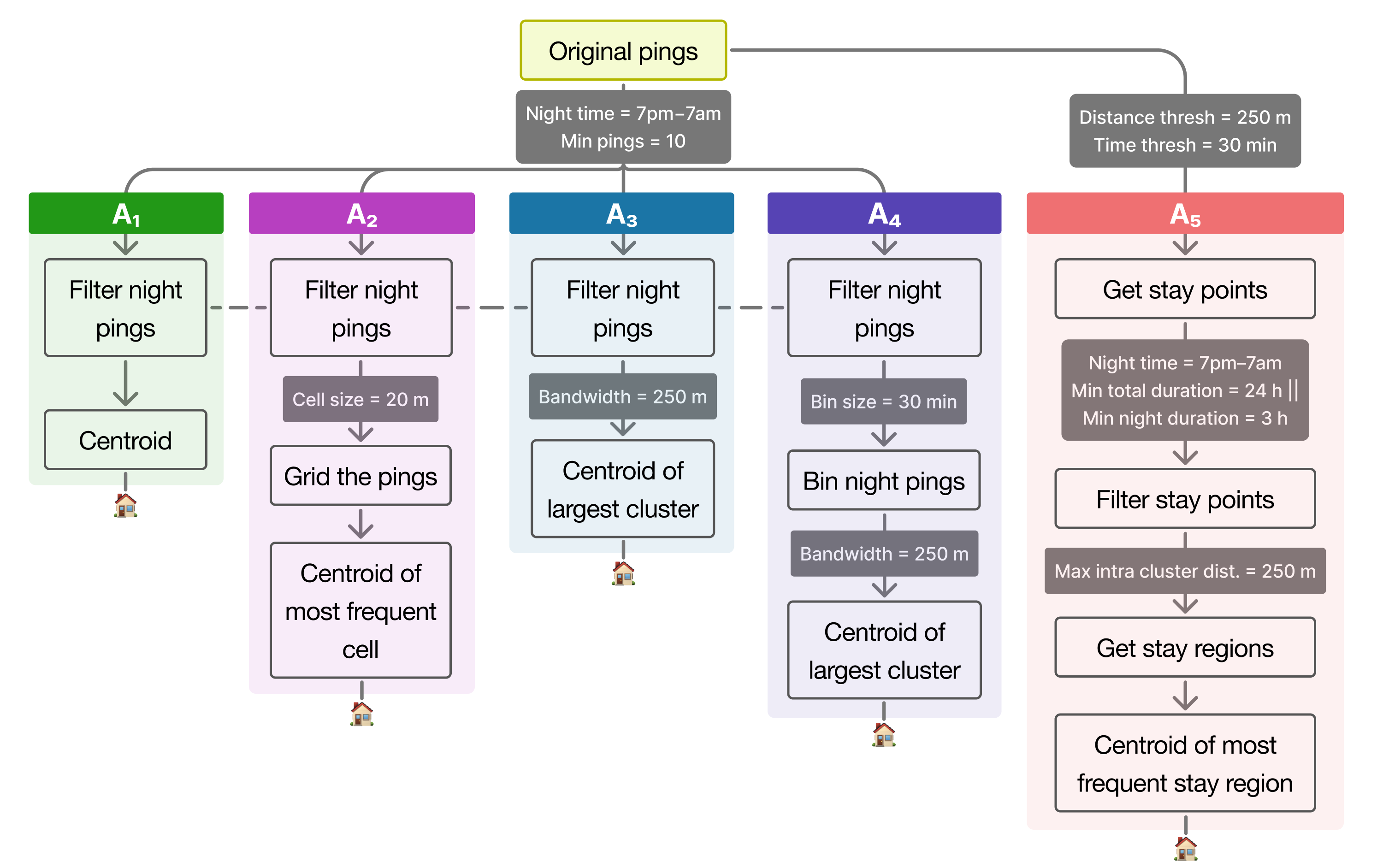}
    \caption{\textbf{Flowchart of the steps of the HDAs compared in this study}. The values shaded in grey depict the algorithms' parameters. The dashed lines between two HDAs depict the same or equivalent step between the two HDAs.}
    \label{fig:hda-flowchart}
\end{figure}

Five HDAs are compared in this study, including a simple baseline algorithm, three algorithms listed in the `Passive GPS data' section of Table \ref{tab:lit-review}, and a derivative of one of those algorithms as proposed in this study.
The steps involved in these algorithms, labeled as $A_1,...,A_5$, are illustrated in Fig.~\ref{fig:hda-flowchart}. The same input dataset is used for each of these HDAs. For clustering-based methods, the implementations of \texttt{scikit-learn}, a popular Python-based machine learning library, are used. The common set of users resulting from each of these HDAs is used for subsequent performance assessment.

\subsubsection{A1: Centroid method}
This is the simplest of all the considered HDAs and is meant to serve as the baseline for comparison with the other algorithms.
In this case, a user's home location is simply computed as the centroid (or alternatively the medoid) of all their nighttime ping locations over the entire study period, following the assumption that a person's most probable location during the night is their home.
This is similar to, but not exactly the same as, most popular cell tower-based algorithms in the case of CDR data \cite{jiang2017activity,chen2018enriching,leca2015significant}.

\subsubsection{A2: Grid frequency method}
This HDA was used in Zhao et al. (2022) \cite{zhao2022estimating}. They first divided the study region into a square grid with cells of 20 x 20 meters.
They considered the home location as the mean location of the pings of the cell with the most nighttime pings over the study duration.

\subsubsection{A3: All-time clustering method} \label{sec:def-a2}
This method involves finding the most popular cluster of all the pings in the nighttime data taken together without distinguishing the temporal variation in locations during this night time.
Though several clustering methods exist as explained in Section \ref{sec:unsupervised-methods}, this method particularly uses mean-shift clustering with the same parameters as in \cite{yabe2020effects, yabe2022mobile, kanasugi2013spatiotemporal}.
All these studies use a flat kernel with a radius of 250 m for KDE. In this study, other parameters in this method such as the sampling strategy for the KDE process and the number of iterations in the hill climb process are controlled to prefer accuracy over runtime speed.

\subsubsection{A4: Binned clustering method}
HDA $A_3$ uses clustering of all the nighttime points at once, meaning it does not distinguish between the following cases: (i) a scenario where most of the nighttime points are concentrated in a small time period (e.g., 10:00 – 10:20 PM) where the user might possibly be in movement and thus more likely to enable LBS, and (ii) a scenario where the same number of nighttime points as in case (i) are distributed evenly across the night.
It can be argued that the latter case provides more confidence in the inferred home location since it relies on better-sampled data.

To overcome this limitation of $A_3$, we propose an adaptation in the form of $A_4$ where the nighttime points are collected at fixed time intervals over the study period. The centroids of these locations are computed and used as inputs in mean-shift clustering.
Similar to $A_3$, the centroid of the largest cluster is labeled the home location.
This HDA introduces a parameter in addition to those of $A_3$ – the binning period, which is taken as 30 minutes in this study.

\subsubsection{A5: Stay-point method}
This HDA was proposed by Sadeghinasr et al. (2019) \cite{Sadeghinasr2019} where they used the stay point detection algorithm proposed by Li et al. (2008) \cite{Li2008} to first identify stay points and then cluster the stay points using hierarchical clustering into stay regions by setting a threshold of a maximum intra-cluster distance of 250 m.
Then, they considered the home locations as the most visited stay regions during nighttime (8 PM – 5 AM) which had a visit duration of at least 3 hours during the nighttime or a total duration of at least 24 hours.

\section{Experimental setup}

\subsection{Data description}
\subsubsection{Smartphone GPS data} \label{sec:gps-data}
This study uses GPS trace data collected using LBS on smartphones and tablets, aggregated and anonymized by a private vendor. The trace table (illustrated in Table \ref{tab:gps-data-sample}) comprises events (called `pings' here) which include a mobile device's (called `user' here) anonymized identifier, latitude and longitude of the point, an estimate of the radius of GPS recording error for that ping, and the Unix-style timestamp of the event (seconds passed since Jan 1, 1970 UTC+00:00).  More details about the data are provided in the Supplementary Section 1.
% , given by the number of seconds passed since UTC+00:00 of Jan 1, 1970.

\begin{table}[hbt!]
\footnotesize
\centering
\caption{A sample of the GPS data used in this study. The coordinates have been fuzzed for illustration.}
\label{tab:gps-data-sample}
\setlength\tabcolsep{3pt}
\begin{tabu} to \textwidth {l X[1.5] X[0.4,c] X[0.35,c] X[0.6,c] X[0.4,c]}
\toprule
\textbf{Row} & \textbf{Device ID} & {\textbf{Longitude}} & {\textbf{Latitude}} & \textbf{Timestamp (s)} & {\textbf{Error radius (m)}} \\
\midrule
1 & 107258c2-c027-41c9-aa4d-166951bd5007 & -86.964964 & 40.064320 & 1552588288.0 & 22 \\
2 & ad96c788-965d-4074-bf28-306a3cf6cb07 & -85.982222                             & 39.848495 & 1552594937.0 & 6 \\
% 3 & 8bee361d-bb1c-4f45-bd4a-4211ac86301a & -85.866489 & 39.615425 & 1552565774.0 & 10 \\
... & … & … & … & … & … \\
... & d3286a43-a68c-42cf-ba71-e838e2276b1a & -86.514245 & 41.672555 & 1552579184.0 & 7 \\ \bottomrule
\end{tabu}
\end{table}

% Data filtering
LBS data is usually slightly erroneous due to inaccuracies in the GPS logging system and thus needs preprocessing for better results. The preliminary data filtering done to create the dataset samples includes removing pings with an error radius of more than 50 m, those with segment speed of more than 50 m/s (180 km/h), and those with acceleration outside the range of -10 to 10 $m/s^2$ (based on works like \cite{nyhan2016predicting,bohm2022gross}). For reference, for the  $i^{th}$ ping in the sequence trace with coordinates $\mathbf{x}_i=(x_i,y_i)$ and timestamp $t_i$, its speed is given by $v_i = \frac{d(\mathbf{x}_{i},\:\mathbf{x}_{i-1})}{t_i - t_{i-1}}$ and the acceleration by $a_i=\frac{v_i - v_{i-1}}{t_i - t_{i-1}}$, where $d$ is the Haversine distance function. By definition, $v_1=a_1=a_2=0$.

% Since the data points contain noise due to recording errors, such points need removed. These include filtering 
% i.e., pings with GPS error radius of more than 50 m were removed. 
% Erroneous points also include points with positions that yield unrealistic derivative values such as speed and acceleration. To ensure the inclusion of more realistic pings, 
% the pings with a speed more than 50 m/s (180 km/h) and acceleration outside of the range [-10, 10] $m/s^2$ were removed \cite{nyhan2016predicting,bohm2022gross}.
% If the $i^{th}$ ping of a GPS trace has coordinates $\mathbf{x}_i=(x_i,y_i)$ and timestamp $t_i$, then its speed is given by $v_i = \frac{d(\mathbf{x}_{i},\:\mathbf{x}_{i-1})}{t_i - t_{i-1}}$ and the acceleration by $a_i=\frac{v_i - v_{i-1}}{t_i - t_{i-1}}$, where $d$ is the Haversine distance function. By definition, $v_1=a_1=a_2=0$.

\subsubsection{Study regions and periods}
\begin{figure}[hbt!]
    \centering
    \includegraphics[width=\textwidth]{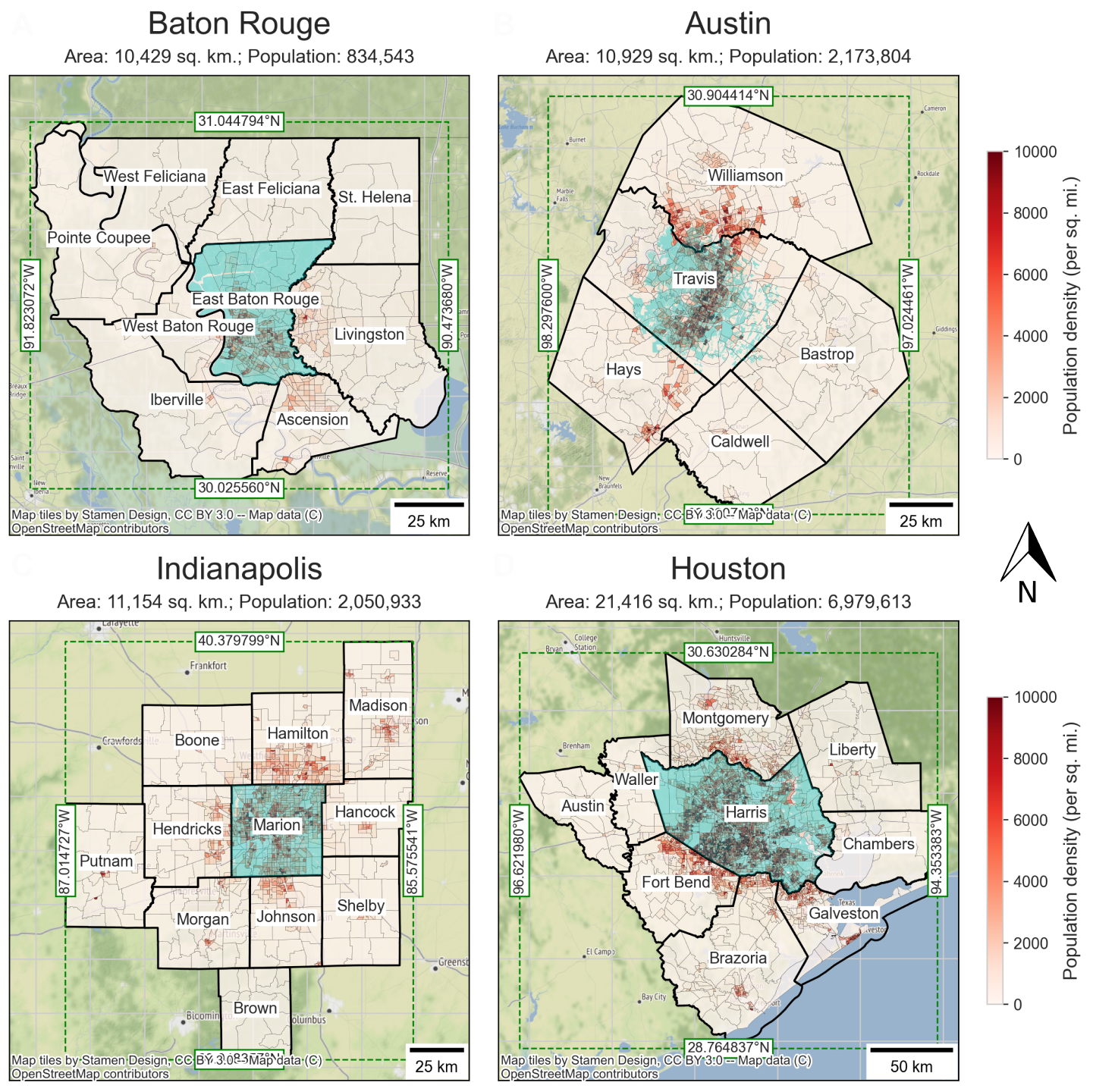}
    \caption{\textbf{Study regions showing the metropolitan statistical area (MSA) counties and bounding boxes}. The population density of the census block groups as per ACS 2020 data is colored in red. The regions covered in the land use maps are shaded in cyan.}
    \label{fig:study-regions}
\end{figure}

Four U.S. metropolitan statistical areas (MSAs) are assessed in this study – (i) Austin, TX, (ii) Baton Rouge, LA, (iii) Houston, TX, and (iv) Indianapolis.
The counties included in these MSAs, their total area, and their total population (as of the 2020 5-year estimates of the American Community Survey (ACS)) are shown in Fig.~\ref{fig:study-regions}.
These regions are chosen from the cities with available land use and smartphone GPS data so as to cover a diverse set of scales and land use patterns. Baton Rouge has a large but sparsely populated MSA, whereas Houston has a much larger MSA. Austin and Indianapolis lie in between but represent cities with very different land use distributions and urban layouts. Houston is known for its sprawling layout, with significant suburban development extending into multiple counties. The city is characterized by a lack of zoning laws, which has led to a unique pattern of residential, commercial, and industrial areas often being interspersed \cite{qian2010zoning}. Austin is characterized by a higher density in the city center, with the urban core being home to a mix of residential, commercial, and cultural facilities (towards mixed-use developments) \cite{handy1998}. By conducting analysis for these four cities with different socio-economic contexts, underlying data characteristics, and scale and complexity, we ensure that our tests are robust and generalizable across various urban settings.

% Datasets
In addition to spatial variation, the datasets used for testing the HDAs are created so as to include temporal variation as well. Particularly, two case studies are chosen to represent the potential temporal difference of HDA outputs before and after two specific events.
The first event is Hurricane Ida which caused damage in southeastern Louisiana upon landfall on August 29, 2020, causing waves of evacuation and displacement around the region, including in Baton Rouge.
% (\textbf{cite})
The periods depicting stability before the landfall, during the mobilization period around landfall, and long after the event are considered in this analysis.

The second event is the first government-mandated lockdown in Indiana on March 16, 2020
% (\textbf{cite})
following the outbreak of COVID-19 in the United States which was known to have drastically reduced mobility.
% (\textbf{cite})
A before-after comparison of the HDAs of these events is deemed useful in explaining the robustness of the HDAs. This is explained in Section \ref{sec:impact-to-apps}.

With these two combinations of study regions and periods, a total of 8 datasets are prepared for testing the HDAs. These are shown in Table \ref{tab:datasets-summary}. The number of unique devices (referred to as `users') obtained after cleaning the GPS data and their ratio to the regional population are also shown. Similarly, the number of filtered pings is also shown for each dataset, reflecting the scale of variation in the test datasets. Note that the pings are filtered within the regions' bounding boxes (shown in green dashed outlines in Fig.~\ref{fig:study-regions}) instead of filtering within the MSA counties for the sake of performance speed.

\begin{table}[hbt!]
\footnotesize
\centering
\caption{Description of the study datasets (combinations of region and analysis period).}
\label{tab:datasets-summary}
\setlength\tabcolsep{4pt}
{\renewcommand{\arraystretch}{0.8}%
\begin{tabu}  to \textwidth {l X[1.0] X[1.5] X[0.6,r] X[0.8,r] X[0.7,r] X[0.8,r]}
\toprule
\textbf{ID} & \textbf{Region} & \textbf{Period} & \textbf{\#(Days)} & \textbf{\#(Users) (k)} & \textbf{\% of Popu.} & \textbf{\#(Pings) (M)} \\
\midrule
% Baton Rouge
$D_1$ & \multirow[t]{3}{2cm}{Baton Rouge, LA} & Aug 1–25, 2021 &	25 & 165.5 & 19.8\%	& 245.8 \\
\cmidrule(l){3-7}
$D_2$ & &  Aug 26 – Sep 7, 2021 & 13 & 87.7 & 10.5\% & 65.5 \\
\cmidrule(l){3-7}
$D_3$ & &  Sep 8 – Nov 30, 2021 & 84 & 316.5 & 37.9\% & 1,120.0 \\
\midrule
% Indianapolis
$D_4$ & \multirow[t]{3}{2cm}{Indianapolis, IN} &  Mar 1–15, 2020 & 15 & 273.9 & 13.4\% & 101.6 \\
\cmidrule(l){3-7}
$D_5$ & &  Mar 16–31, 2020 & 16 & 251.6 & 12.3\% & 127.2  \\
\cmidrule(l){3-7}
$D_6$ & &  Mar 1–31, 2020 & 31 & 445.1 & 21.7\% & 241.6  \\
\midrule
% Austin
$D_7$ & Austin, TX &  Jul 1–7, 2021 & 7 & 166.9 & 7.7\% & 97.6 \\
\midrule
% Houston
$D_8$ & Houston, TX &  Jul 1–7, 2021 & 7 & 538.8 & 7.7\% & 331.1 \\
\bottomrule
\end{tabu}
}
\end{table}

\subsection{Performance metrics}\label{sec:metrics}
In the absence of ground-truth information on device users' home locations, the accuracy of the HDAs is tested using three approximate or pseudo-performance metrics. All these are based on some assumptions that are generally considered valid intuitively and in the literature.

\subsubsection{M1: Residential detection rate}
% Definition
This metric makes use of the idea that a good HDA should detect more homes in a city's residential areas as opposed to other land use categories such as commercial, industrial, and forests.
This metric is given by the proportion of homes detected by a given HDA in the residential area of the region based on its land use distribution (see Supplementary Section 2.1 for more details).
To offset some potential mislocation errors due to the nature of the GPS data and the often convoluted land use maps, tolerance buffers of different widths are also considered in the calculation. This results in the following definition of the performance metric:

\begin{equation}
    M_1(A) = \sum_{r=0}^{r_{\text{max}}}
    w(r)\rho_{A}(r) \text{, where }
    w(r)=\frac{r_{\text{max}} - r}{\sum_{r=0}^{r_{\text{max}}}r_{\text{max}} - r}
    \label{eq:m1-definition}
\end{equation}

Here, for buffers of width $r$, ranging from zero to $r_\text{max}$, $\rho_A(r)$ is the proportion of homes detected in the combined buffered residential area detected by HDA $A$. For instance, a value of $M_1(A)=0.4$ can be roughly interpreted as 40\% of the users' home locations detected by HDA $A$ lying within a region of the city classified as `residential'.
In the subsequent experiments, the value of $r_{\text{max}}$ is taken as 50 m, with buffer width increments of 5 m, the same as the maximum allowed error in GPS spatial accuracy as explained in section \ref{sec:gps-data}.

\subsubsection{M2: Proximity to daily data}
\label{sec:m2-definition}
% Definition
This metric uses the idea that a home location should be the origin/destination of one’s daily trips. Given a user's home location detected by a given HDA, this metric involves computing its distance to the closest ping in that user's nighttime pings on each day in the study period using Haversine distance. Then, the median of these daily shortest distances is taken for each user.
The cumulative density function (CDF) of this median shortest distance is drawn and the normalized area under the curve is computed. This represents the proximity performance metric, given by the following:
\begin{equation}
    M_2(A) = \frac{1}{\delta_{\text{max}}} \int_{\delta=0}^{\delta_{\text{max}}} F_{A}^{\Delta}(\delta)\cdot d\delta,\;\;\;\;\text{where}\;\; \delta_{A,i}=q^{0.50}_{t\in 1:n_T}\left(\min_{x_{i,t}\in \mathbf{X}_{i,t}} \|h_{A,i} - x_{i,t}\|_2 \right)
    \label{eq:m2-definition}
\end{equation}

Here, $F_A^{\Delta}$ is the CDF of the median shortest distance of the users detected by HDA $A$, $\delta_{A,i}$ is the median shortest distance for the user $i$ whose home location is given by $h_{A,i}$,  $\mathbf{X}_{i,t}$ is the set of nighttime pings at night $t$, $q^{\text{0.50}}$ represents the 0.50 quantile (that is, median) over all study days up to $n_T$, and $\delta_{max}$ is a reasonable upper limit, taken as 5 km.

\subsubsection{M3: Home stay duration}
% \subsubsection{M3: Time spent at visited locations}

% This metric is based on the notion that, during nighttime hours, an individual's home is typically their most frequented destination. By analyzing data collected over multiple days, this metric involves calculating the frequency with which a person visits each location during nighttime hours. The resulting information is then used to compare the individual's most frequently visited location with their detected home location.

This metric is based on the idea that people typically spend the majority of their nighttime at their homes. For a given user, we first identify the locations they visit during nighttime hours using a stay region detection method similar to Sadeghinasr et al. (2019) \cite{sadeghinasr2019estimating} but with an adaptive linkage calculation (details of this method are provided in Supplementary Section 2.2).
The stay region closest to each user's detected home location is assigned as their `home region'.
With the detected stay regions, the performance metric for each user is simply the ratio of time outside the home region to the total time spent in all stay regions. The overall performance metric is given by the area under the curve of the CDF of this value:

\begin{equation}
    M_3(A) = \int_0^1 F_A^{\tau}(r)\cdot dr,\;\;\;\;\text{where}\;\;r_{A,i}=\frac{\tau(C_{h_{A,i}})}{\sum_{k=1}^{K_i} \tau(C_{h_{k,i}})}
    \label{eq:m3-definition}
\end{equation}

Here, $F_A^{\tau}$ is the CDF of the ratio, $r_{A,i}$, of time ($\tau$) spent in the home stay region, $C_{h_{A,i}}$, to the maximum time spent in any stay region $C_{k,i}$ over all the users $i$ detected by HDA $A$, and $K_i$ is the total number of stay regions detected for user $i$.
Similar to $M_1$ and $M_2$, a higher value of $M_3$ indicates a better HDA.
% Note that the inverse of the CDF is simply used to make it consistent with the interpretation of the other two performance metrics and does not differ from $F_A^{\tau}$ for comparing the HDAs. This means that similar to $M_1$ and $M_2$, a higher value of $M_3$ indicates a better HDA.

% To evaluate an HDA's effectiveness, the proportion of time an individual spends at their home location is compared to the maximum time they spend at any location during nighttime hours. Ideally, this ratio should be equal to 1 for all users and lower values of this ratio for large number of users is undesired. This ratio is computed for all users ($i$) across all algorithms ($A$) as follows:

% \begin{equation}
%     r_{A,i} = \frac{t(H_{A,i})}{\max_k t(C_{ki})}
%     \label{eq:m3_ratio}
% \end{equation}

% We first evaluate the cumulative probability distribution of the ratio across the users for all algorithms. The overall performance metric for evaluating each algorithm, in this case is calculated

% \begin{equation}
%     m_3(A) = 1 - 
%     \label{eq:m3}
% \end{equation}

\section{Results}
The HDAs listed in Section \ref{sec:algorithms} are compared on the basis of their precision, as approximated by the three performance metrics in Section \ref{sec:metrics} and their sensitivity to data quality. These are described in the subsequent sections.

% Key things to ask
% \begin{enumerate}
%     \item What did we want to achieve in this study?
%     \item What methods and performance metrics did we use? How are they justified?
%     \item Which method works out the best and in what cases?
%     \item How sensitive are our results w.r.t. the algorithms’ parameters?
% \end{enumerate}

\subsection{Performance comparison}\label{sec:performance-comparison}

The visual comparison of the performance metrics $M_1$, $M_2$, and $M_3$ across the HDAs over all the datasets is shown in Fig.~\ref{fig:performance-radar}. The generating curves of these metrics are provided in Supplementary Figure 2. In Fig.~\ref{fig:performance-radar}, the size of the radar polygons depicts the overall performance of an HDA, while the skewness of the polygons hints at the differences in the behavior of the HDAs across different datasets.
\begin{figure}[!hbt]
    \centering
    \includegraphics[width=\textwidth]{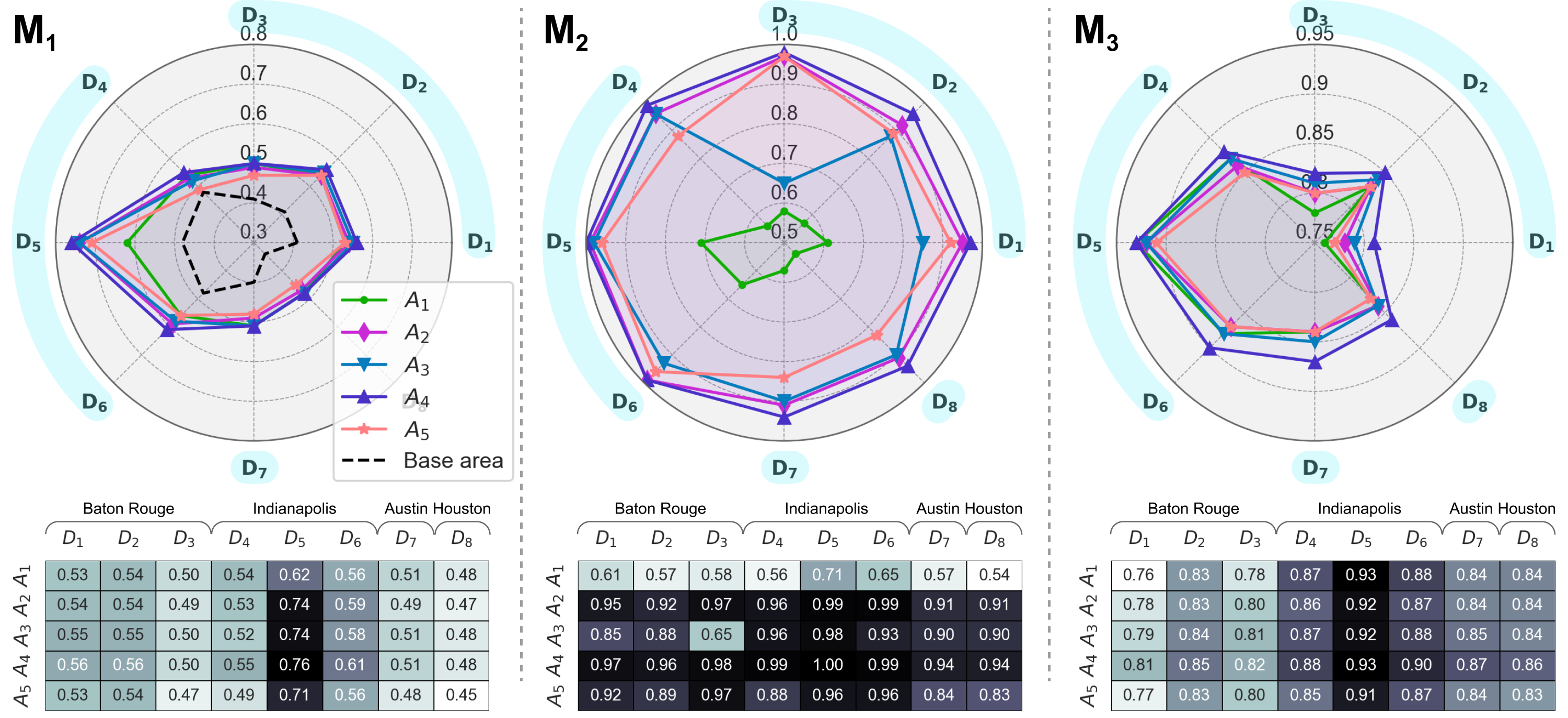}
    \caption{\textbf{Performance metrics for the HDAs across the study datasets}. For $M_1$, the dashed black line represents a uniform random selection algorithm based on the residential area buffers up to 50 m. The datasets of the same city are grouped in cyan.}
    \label{fig:performance-radar}
\end{figure}
\subsubsection{Overall differences by HDA}
% worst  of $A_1$
The findings from the plots in Fig.~\ref{fig:performance-radar} are diverse and vital.
First, $A_1$ consistently performs the worst in these overall results. This is expected, as $A_1$ is a very straightforward HDA with several key limitations: (i) It is difficult to find the most frequent place among GPS points is not easy due to high data precision; (ii) The centroid may not necessarily be the most probable location; (iii) The results of this method are heavily susceptible to disturbances due to outliers; (iv) This method does not distinguish between spatiotemporal regions of stay and movement.
For people with high movement during the night, the mean value of the coordinates can shift the detected home substantially far away from the user's trajectory. This explains why $A_1$ performs substantially worse in the case of $M_2$ compared to the other HDAs, since $M_2$ directly involves computing the distance of the detected home location with the closest nighttime trajectory point.

% best performance of $A_4$
The performance of the other algorithms is largely similar, with some exceptions. Algorithm $A_4$ consistently performs better than the others, as is evident from the largest radar polygons corresponding to $A_4$ in the three metrics. In particular, although $A_4$ requires a data filtering criterion on its base HDA $A_3$ and thus operates on fewer data points than $A_3$, it performs better than that. This might be attributed to the focus on data quality over quantity by discretizing the data temporally, as explained in Section \ref{sec:def-a2}. This is important because it is possible for users to have high LBS activity during traveling (e.g., for navigation services) which may overshadow the location data during stay periods such as at home. Since traveling generally occurs far from home, all HDAs other than $A_4$ are more likely to consider these irrelevant points for the home detection process.

% other HDAs: $A_2$
This bias is reduced to a lesser extent in $A_3$ and $A_5$ that rely on clustering.
This positive impact of discretization is also evident in terms of space. $A_2$, which is a very simple heuristic that only involves finding the most visited grid cell, i.e., the discretization of space, performs \sout{remarkably well, with metric values finishing}, with metric values finishing close to $A_4$ in most cases.

% other HDAs: $A_5$
The rule-based HDA $A_5$ is generally found to perform slightly worse than $A_3$, although this pattern reverses in the case of $M_2$. Both $A_3$ and $A_5$ involve clustering, but the order and kind of clustering are different between the two. It may be argued that the time and distance-based thresholds involved in the stay point detection step of $A_5$ might hamper the performance of the algorithm since those thresholds do not take into account the continuity of the data.
% In $A_5$, the stay points of each user's trajectory are first computed based on thresholds on distance and time between ping sequences, data filters pertaining to nighttime are then applied, and finally, the stay regions are obtained using hierarchical clustering.

\subsubsection{Differences by dataset}
% effect of COVID-19 lockdown (D4, D5 & D6)
The radar plot in Fig.~\ref{fig:performance-radar} also shows the significant differences in the performance of the same HDA in different datasets. Notably, all the metrics are observed to be the highest in the case of $D_5$. It should be noted that datasets $D_4$ , $D_5$, and $D_6$ have the same underlying urban land use and transportation networks. $D_5$ corresponds to the period of reduced mobility and high stay-at-home rates during the surge of COVID-19 in the Indianapolis region. It includes the date of the first death related to COVID-19 recorded in the region on March 16, 2020, and the imposition of the government-mandated lockdown on March 23 \cite{indylockdown2020}. Since people were more likely to stay at home during the period of $D_5$, the data quality for the HDAs was substantially better than the other datasets, making it easier for all the HDAs to perform the best. This is made further prominent in the stark difference between $D_4$ and $D_5$ in the value of $M_3$ which depends on the time spent at home.

Moreover, the performance metric values for $D_6$ are consistently near the corresponding values of $D_4$ and $D_5$. This makes sense given that the period of $D_6$ is the union of the periods of $D_4$ and $D_5$ which are of equal length. This indicates that the better data quality of $D_5$ does not inordinately skew the performance metrics.%, implying that these metrics are not overly sensitive to.
% This also serves as a basis to compare the consistency of the HDAs since a good HDA should be expected to perform well on both $D_4$ and $D_5$, and by extension $D_6$, which is the union of the periods assessed in $D_4$ and $D_5$. It can

% effect of hurricane in Baton Rouge
In Baton Rouge, the effect of Hurricane Ian is observed to be small yet important. This is evident in the higher values of $M_3$ for dataset $D_2$ that corresponds to the period close and immediately after the hurricane landfall compared to the pre-landfall ($D_1$) and long-term post-landfall ($D_3$) periods. However, the values of $M_1$ and $M_2$ do not vary significantly between $D_1$, $D_2$, and $D_3$.

\subsubsection{Differences by metric}
The ranges and behaviors of the three performance metrics also shed light on the nature of the analysis of this study. First, $M_1$ has a large range of 0.45 to 0.76. All the tested HDAs perform substantially better than a random uniform HDA where the residential detection rate curve is plotted by simply computing the proportion of land use region covered by residential areas. This is evident in Fig.~\ref{fig:performance-radar}A where the black dashed curve (denoting this uniform random HDA) is significantly smaller than those of the other HDAs in the plot.
It must be noted, however, that $M_1$ relies on assumptions about home location that might not always hold true and could have skewed the results. For example, some users may stay at places other than their homes (such as a hotel or a relative's residence). Similarly, the home locations of night-shift workers may be overrepresented in the commercial areas of a city and thus reduce the value of $M_1$.

The case for $M_2$ is also similar. It has a substantially small range outside of the poorest performing $A_1$. This could be attributed to the fact that $M_2$ is unidirectional in its utility. That is, a small shortest distance of trajectory points from home only serves as a necessary condition for a good HDA, not a sufficient condition. Its computation relies on the distance to the closest point to the trajectory. Since the home locations are detected based on the trajectory itself, it is highly probable for an HDA to produce a high value of $M_2$ for a set of users who do not travel very long distances.

\subsection{Sensitivity to data quality}\label{sec:sensitivity}
% Why we need this exercise
In the previous sections, we observed the difference in the performance of the test HDAs. While it was shown that the continuity of data discretized in space and time substantially influences the goodness of an HDA, there is substantial nuance to the effect of data quality in terms of overall ping density on this goodness. In this section, we particularly ask the question: ``if an analyst is interested in using any of the tested HDAs for their mobility analysis purpose, how should the choice of the HDA be based on the quality of the data available to them?"
``if an analyst has geolocation data of a specific ping density, which HDA should they choose for their analysis?"

% What are we doing
Building on the notion of ping density, the data quality of a user in this section is defined as the mean number of pings per night in their data points. Users with more pings on average are expected to have higher quality data and yield better home location detection results. At the same time, however, owing to the nature of mobile phone geolocation data, most users have very few data points, making home detection a difficult task (for reference, see SM). This means that a good HDA should strike the balance between good data quantity and quality.

\begin{figure}[!hbt]
    \centering
    \includegraphics[width=\textwidth]{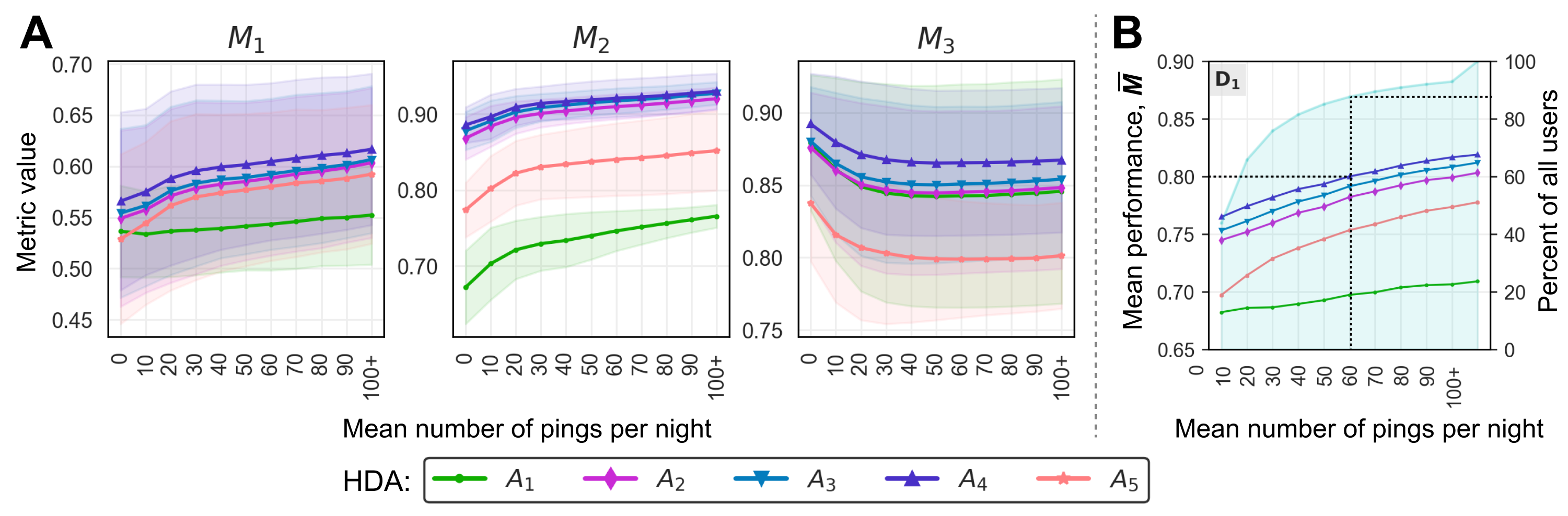}
    \caption{\textbf{Impact of data quality on HDA performance}. Each value $x$ on the x-axis represents the subset of users having at least $x$ pings per night on average. (A) Comparison of the mean value ($\bar{x}$) of each metric across all the datasets. The shaded regions correspond to the range $\bar{x}\pm \sigma$, where $\sigma$ is the standard deviation across the datasets (B) Comparison of the mean of the three metrics for one dataset. For reference, the CDF of the users sorted by the average nightly ping count (x-axis) is shown in the shaded blue curve on the right y-axis.}
    \label{fig:sensitivity-analysis-main}
\end{figure}

% How do we do this
To achieve this, we recomputed the performance of the HDAs on several subsets of the users by dividing them by their data quality, given by their mean nightly ping count. To simplify the decision-making for HDA choice, we further computed the mean value of the three metrics for the subset of users contained in each bin, given by $\bar{M}=\frac{1}{3}(M_1+M_2+M_3)$. The results of these aggregate metrics are shown in Fig.~\ref{fig:sensitivity-analysis-main}.

% How do we interpret this
The findings of this figure are aligned with those in the previous section.
First, we see here that at nearly all levels of data quality, the order of performance is largely consistent with the overall results shown in Fig.~\ref{fig:performance-radar}. $A_4$ still consistently performs the best, closely followed by $A_3$ and $A_2$, while $A_1$ and $A_5$ perform considerably worse.
When the data quality is measured in relative terms, i.e., using the ping count distribution of each dataset, the trends are considerably different (see the Supplementary Figure 3).

% Panel A: By metric
Notably, in Fig.~\ref{fig:sensitivity-analysis-main}A, though $A_1$ performs worse than $A_5$ in the case of $M_1$ and $M_2$, the trend is reversed for $M_3$. The trends of $M_3$ are also different from those of the other two metrics in that, unlike them, $M_3$ decreases with increasing data quality. It is likely because it involves computing the ratio of time spent in the detected stay-at-home region, which is likely to be exactly the same as the only (or one of the only) stay region detected for low-quality data users since they do not have enough data, to begin with. In contrast, $M_1$ and $M_2$ rely on the richness of the data in increasing the likelihood of locating a user in a residential region and closeness to the trajectory respectively.

% Panel B: By dataset
To compare the overall relative performance of the HDAs, we also computed the mean value of each of the three metrics across all the study datasets. The result of one dataset $D_1$ is shown in Fig.~\ref{fig:sensitivity-analysis-main}B. Similar results for the other datasets are shown in the Supplementary Figure 4.
It can be seen that the opposite trends of $M_1$ and $M_2$ with $M_3$ are balanced to some extent when their values are averaged. There is a steady but small increase variation in the value of $\bar{M}$ as the user  quality increases in $D_1$. This shows that there is merit in choosing these metrics as their values do not show any abrupt behavior over different data quality categories.

% use of this figure for researchers
This comparison is also helpful in making the choice of data filtering required for any downstream application of home location detection. For example, suppose we decide that a mean performance value of 0.8 is acceptable in a dataset similar in ping count distribution to $D_1$ and an urban land use similar to Baton Rouge. Then, we can refer to Fig.~\ref{fig:sensitivity-analysis-main}B to see that, for example, for HDA $A_4$, users with at least 50 pings per night would be required for analysis (dotted vertical line). This corresponds to the 13\% best quality users of the dataset since 87\% of the users have fewer than 50 pings (right horizontal dotted line).

% decisions based on data quality available to researchers. For example, by setting a threshold of an acceptable level of performance, such as $\bar{M}_{\text{max}}=0.7$.
% LZ: From our last discussion, two points can be raised here: 1. Our metrics are consistent under different data quality; 2. By combining three metrics using equation ..., we find the combined values are stable. This can be useful in deciding an absolute threshold for HDAs.

\subsection{Impact on applications}\label{sec:impact-to-apps}
To see how different HDAs would influence applications of human mobility assessment and how our performance metrics could help improve the results, we conduct two experiments on common tasks where smartphone data is considered superior to other sources.
These are explained in the subsequent sections.

\subsubsection{Hurricane evacuation identification}
% Short description
Large-scale GPS data is used to estimate the evacuation/return patterns during natural disasters~\cite{yabe2019cross, zhao2022estimating}. In this task, a crucial factor is the distance between individuals' post-disaster stay locations and original home locations before the disaster.

% How the experiment is conducted
Here, we calculate this factor based on $D_1$ (before landfall) and $D_3$ (aftermath of Hurricane Ida) using the five test HDAs. Then, we estimate the evacuation ratio using the threshold, i.e., if the distance between an individual's pre- and post-disaster home locations exceeds 1 km, we consider them as evacuated.

\begin{figure}[!htb]
    \centering
    \includegraphics[width=\linewidth]{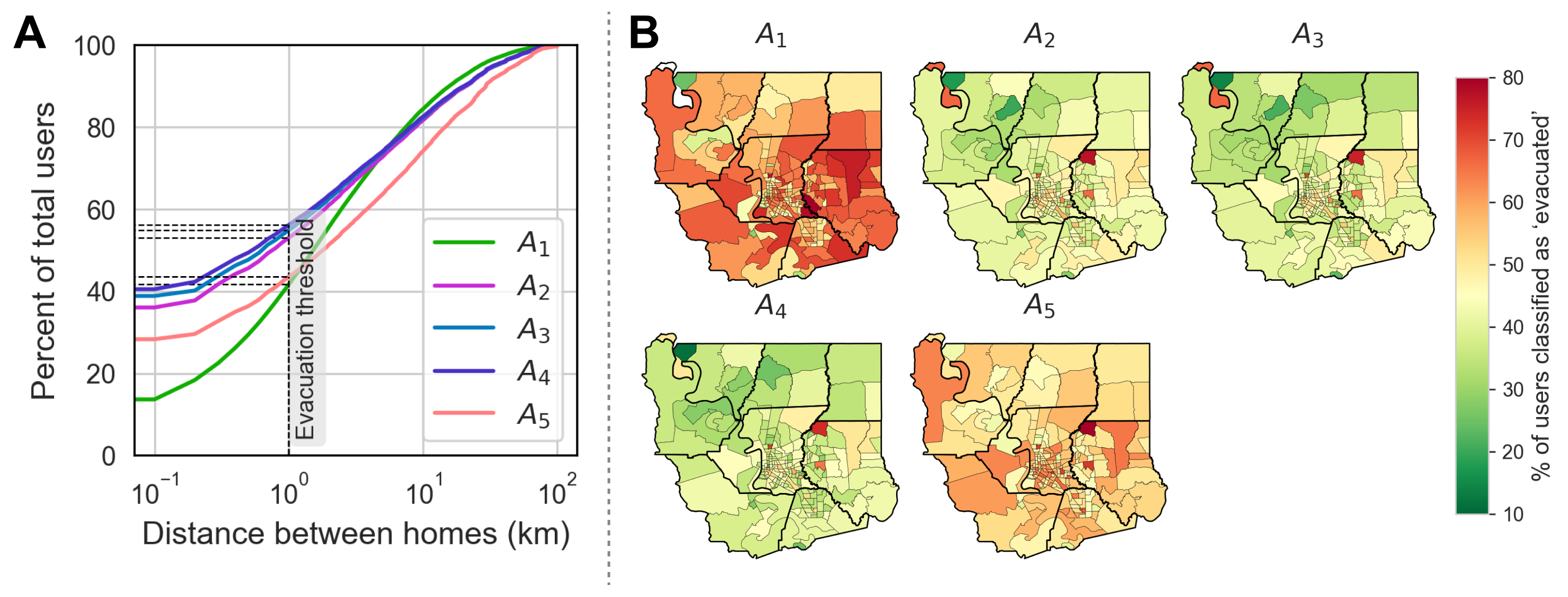}
    \caption{
\textbf{Evacuation identification results under different HDAs}. (A) CDF of the distance between homes identified before (dataset $D_1$) and after Hurricane Ida ($D_3$) in Baton Rouge MSA. The threshold to classify users as displaced (1 km) is highlighted. (B) Identified percentage of users classified as evacuated by census tracts.}
    \label{fig:evacuation-result}
\end{figure}

% Results
%% To Shagun: can your results strengthen these points, if so, I will leave this paragraph like this.
We observe that among the five HDAs, $A_1$ and $A_5$ produce significantly different distributions of the distance between pre- and post-disaster homes (see Fig.~\ref{fig:evacuation-result}.). Even for HDAs with similar CDF curves, it can be seen from Fig.~\ref{fig:evacuation-result}B that they can generate a significant estimation of evacuation ratio in some areas (e.g., the northwestern part and the southern part of the city). When connecting these results with the observations of Section~\ref{sec:performance-comparison}, we notice that the HDAs with good and similar performance metrics (namely $A_3$, $A_4$, and $A_2$) tend to create similar results. In contrast, $A_1$ and $A_5$ result in much higher evacuation rates. Since evacuation rates are essential in assessing policies and equity issues related to home evacuation, in-place sheltering, and disaster recovery, it can be imaged that adopting an arbitrary HDA can yield substantial negative impacts on policymaking  \cite{martin2017leveraging}.

\subsubsection{COVID-19 policy impacts assessment}

GPS-based cell phone location data has been extensively used to evaluate mobility patterns and potential solutions during COVID-19 \cite{grantz2020use}. These include evaluating alterations in population-wide mobility \cite{xiong2020mobile}, compliance with COVID-19 policies in various demographic groups \cite{bargain2020trust}, and the spread and associated risk of disease from different regions \cite{verma2021spatiotemporal}. However, erroneous home location inference may lead to inaccurate assessment of mobility changes and policy compliance of regions or demographics, resulting in resource misallocation and ineffective policies.

% How the experiment is conducted
To test this concern, we compare the locations of homes inferred from each HDA in two periods: the pre-COVID-19 normal mobility period (1-15 March 2020; data $D_4$) and the post-lockdown mobility period (16-31 March 2020; data $D_5$). Ideally, high proximity between the homes inferred from both datasets for each HDA should be expected. However, significant inconsistencies are observed in certain HDAs that may lead to inaccurate inferences.

\begin{figure}[ht]
    \centering
    \includegraphics[width=\textwidth]{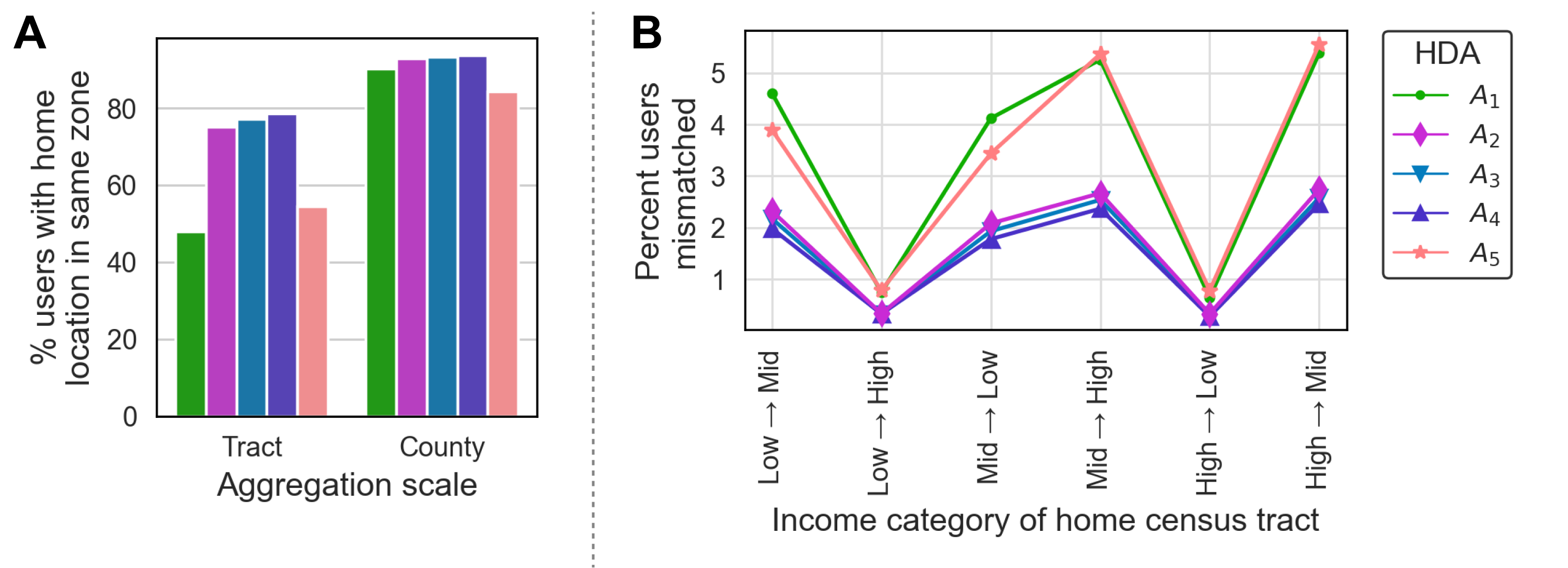}
    \caption{\textbf{Consistency of home locations in the Indianapolis region before ($D_4$) and after ($D_5$) COVID-19 mobility restrictions}. (A)  Percentage of users with consistent home inference in the same zone displayed across different aggregation levels. (B) Percentage of users with income category mismatches in the inferred home tract. `Low $\rightarrow$ Mid' indicates a change from `Low' income category in the restricted mobility period (data $D_5$) to `Medium' category in the pre-COVID normalcy mobility period (data $D_4$).}
    \label{fig:covid19-zonal-match}
\end{figure}

%Results
To demonstrate the consistency of inferred homes, we report the percentage of users with home locations within the same zone for each HDA (Fig. \ref{fig:covid19-zonal-match}A). We show the results for both an aggregated administrated boundary (county) and disaggregated one (tract). With HDA $A_1$, only 47\% of the users exhibit a consistent census tract, while with $A_5$, 56\% of such users were observed. For all remaining users, demographic considerations can be inconsistent and imprecise. For every HDA at the spatially aggregated county level, more than 80\% of the users are classified within the same county. At both spatial levels, $A_4$ shows the highest consistency, with $A_3$ and $A_2$ being comparable. Therefore, for reliable analysis, this suggests using HDAs $A_2$, $A_3$, and $A_4$ rather than $A_1$ and $A_5$, aligning with the findings presented in Section~\ref{sec:performance-comparison}.

We further investigate its potential impact on realistic applications, and income-based inequality assessment, which rely on demographic information inferred from home location.  Income-based inequalities have been extensively examined using cell phone data in aspects such as access to opportunities \cite{mittal2023linking}, the well-being of individuals \cite{pappalardo2016analytical}, emissions \cite{guo2020more}, and evacuation \cite{yabe2020effects}. An inadequate HDA may result in the misclassification of users into different income groups, compromising the accuracy of assessing inequalities and characteristics associated with people from specific income groups.

We assess the percentage of users exhibiting income group discrepancies based on the median income of inferred home's census tract for two datasets for an HDA. Income groups are categorized from the Longitudinal Employer-Household Dynamics (LEHD) Origin-Destination Employment Series (LODES) dataset, comprising three categories based on monthly income: low (less than \$1,250), mid (\$1,250 - \$3,333), and high (\$3,333 and above) \cite{us2020lehd}. Fig. \ref{fig:covid19-zonal-match}B shows the percentage of users with inconsistent income group classification across the two datasets. A minimal proportion of users experienced misclassification between high and low-income groups. However, a significant number of low-income users were incorrectly classified as middle-income and vice versa, resulting in a blending of categories and inaccurate assessment of behavior. Both $A_1$ and $A_5$ exhibit the highest percentage of misclassified users. The consistent performance of $A_2$, $A_3$, and $A_4$ suggests their suitability for studies involving demographics. These findings underscore the importance and precision of the inferred metrics, as these findings align with the results from Section \ref{sec:performance-comparison}.

\section{Discussion and conclusion}
% What we did
In this study, we examine several home detection algorithms (HDAs) for mobile phone geolocation data, an important source that opens novel opportunities on several crucial topics. To evaluate the quality of identified home locations, we propose three performance metrics. Each metric corresponds to a feature that the true home location would likely hold: most identified homes should be located in residential areas (metric $M_1$), the home should be close to one's daily trajectories for every day ($M_2$), and people typically spend most of the nighttime at their homes ($M_3$). We test four representative HDAs together with one which we propose and calculate the metrics on eight datasets in four US cities with different urban layouts and population distributions. We also conduct a sensitivity analysis against data density to understand the impact of data quality on the relative quality of the detected home locations.

We find that different HDAs, even well-established in the literature, can lead to significantly different home location results. Among the five HDAs tested in this study, we observe that two of them ($A_1$ and $A_5$) consistently perform worse than other algorithms in all eight datasets. More than 20\% of the homes detected by these two HDAs fall outside a 2-mile radius from the home locations estimated by the other three HDAs in the eight datasets. $A_1$ is a simple centroid-based algorithm that is primarily used in call detailed records (CDR) mobile phone data. Its poor performance can be attributed to its sensitivity to outlier records and a lack of consideration for other data filtering criteria and nuances. $A_5$ is a more sophisticated algorithm that uses both clustering and a rule-based approach to identify the location of the home. The choice of its many parameters might be attributed to some or all of its poorer performance.
The other three HDAs ($A_3$, $A_4$, and $A_2$) perform similarly to each other. In addition to this, it is found that all three metrics agree with each other in terms of the rank of the performance, which supports the strength of their design.

We also propose a new algorithm ($A_4$) is based on $A_3$ with an additional process to bin every 30-minute pings to consider spatial data continuity. Under our metrics, we report that $A_4$ consistently performs better than other HDAs studied. It is worth noting that by adding the binning process, we also manage to reduce computational time when compared with $A_3$. Although computational time is not a big concern for this offline task, it becomes important if the size of the samples is substantially large.

We a sensitivity analysis of the data quality to provide useful suggestions to researchers who might encounter different data collection frequencies and sample rates. It is found that the order of relative performance remains largely the same even for different subsets of mobile phone devices ranked by their data quality.

% Value proposition
We expect our work can provide the following values to researchers and practitioners who are using HDAs. First, we hope that this study can shed light on a previously unexamined issue: the quality of detected homes and their potential influences on findings in can provide useful guidance for their methodology design.
to facilitate access to our proposed metrics and different HDAs.

% Future directions
We also recognize some limitations of our study and some related topics that merit further examination. First, due to the absence of large-scale true home locations, our evaluation can only be indirect. Note this is also the motivation for performing home detection, which suggests that this would be a limitation for all HDAs when they are applied in practice. Here, we introduce the COVID-19 scenario to alleviate this issue as the impact of the lockdown influence on human mobility is well studied and accepted. Given that the information about people's exact home locations is very sensitive, we expect the restrictions to be unlikely to be fully resolved, but we expect future events to provide opportunities to create more evidence. 
Second, we recognize that the datasets used in this study may not reflect the nature, quality, and quantity of data available to other researchers.
Finally, our proposed metrics are `necessary' conditions in the sense that the detected homes are good, as they align with our intuition of the features that a real home location would follow. It would be interesting to establish the `sufficient' conditions for an HDA's results to be acceptable.
To establish such standards, we posit the need for more and diverse empirical evidence with our proposed metrics.

\section*{List of abbreviations}
\begin{tabular}{l l}
    ACS & American Community Survey \\
    CDF & Cumulative Density Function \\
    CDR & Call Detailed Records \\
    COVID-19 & Coronavirus Disease 2019 \\
    DBSCAN & Density-Based Spatial Clustering of Applications with Noise \\
    GPS & Global Positioning System \\
    HDA & Home Detection Algorithm \\
    KDE & Kernel Density Estimation \\
    LBS & Location-Based Services \\
    LEHD & Longitudinal Employer-Household Dynamics \\
    LODES & LEHD Origin-Destination Employment Statistics \\
    MSA & Metropolitan Statistical Area \\
    UTC & Coordinated Universal Time \\
\end{tabular}

\section*{Declarations}

\subsubsection*{Acknowledgments}
The geolocation data was purchased from a well-established data vendor.
% We thank Quadrant, a subsidiary of Appen Limited, for providing us with the unprocessed smartphone GPS data used in this study under a data purchase contract.

\subsection*{Author contributions}
RV, SM, XC, and ZL conceived the study design. XC reviewed the research literature. SU collected the mobile phone geolocation data and supervised the study. RV, ZL, and SM designed the three performance metrics $M_1$, $M_2$, and $M_3$ respectively. RV analyzed the results and prepared the results for the `Performance comparison' section. ZL and SM prepared the results for the `Impact applications' section. All authors read and approved the final manuscript.

\subsection*{Availability of data and materials}
The complete smartphone geolocation data analyzed in the current study are not publicly available due to a contract with the private data provider. However, samples of the processed datasets are available from the corresponding author upon reasonable request.
% The geographic information systems data layers 
All the relevant code used to process the data sets used in this study is available at \href{https://github.com/rvanxer/home_detection}{https://github.com/rvanxer/home\_detection}.

\subsection*{Funding}
This study was not sponsored by any funding agency.

\subsection*{Competing interests}
The authors declare that they have no competing interests.

\subsection*{Additional information}
A Supplementary Material document is available for this manuscript.

\bibliography{main}

\end{document}